\documentclass[11pt,a4paper]{article}

 \newcommand{\be}{\begin{equation}}
 \newcommand{\ee}{\end{equation}}
 \newcommand{\bl}{\begin{equation}\begin{array}{ll}}
 \newcommand{\el}{\end{array}\end{equation}}
 \newcommand{\bll}{\begin{equation}\begin{array}{lll}}
 \newcommand{\bdm}{\begin{displaymath}}
 \newcommand{\edm}{\end{displaymath}}
 \def\bea{\begin{eqnarray}}
 \def\eea{\end{eqnarray}}
 \def\barr{\begin{array}}
 \def\earr{\end{array}}
 \newcommand{\bean}{\begin{eqnarray}}
 \newcommand{\eean}{\end{eqnarray}}
\def\p{\partial}

\def\f{\varphi}
\def\ve{\varepsilon}
\def\ep{\epsilon}
 
 \def\la{\lambda}
 \def\La{\Lambda}
 \def\al{\alpha}
 
 \def\om{\omega}

\def\half{\frac{1}{2}}

\def\2third{\frac{2}{3}}
\def\4third{\frac{4}{3}}
\def\3quart{\frac{3}{4}}

\def\lim{\rightarrow}

\def\pr{\prime}

\def\bl{\bar{l}}

\def\bl{\bar{l}}

\def\bX{\bar{X}}

\def\btau{\bar{\tau}}

\def\cL{{\cal L}}

\def\da{\dot{a}}
\def\dop{\dot{p}}
\def\dq{\dot{q}}

\def\dvphi{\dot{\f}}

\def\dchi{\dot{\chi}}
\def\dxi{\dot{\xi}}

\def\dpsi{\dot{\psi}}
\def\dal{\dot{\alpha}}

\def\deta{\dot{\eta}}

\def\dF{\dot{F}}

\def\dh{\dot{h}}

\def\hU{\hat{U}}

\def\Lag{{\cal L}}

\def\prpsi{\psi^{\prime}}

\def\prPhi{\Phi^{\prime}}
\def\prchi{\chi^{\prime}}
\def\prho{\rho^{\prime}}
\def\preta{\eta^{\prime}}

\def\prH{H^{\prime}}
\def\prG{G^{\prime}}
\def\prF{F^{\prime}}

\def\tc{\tilde{c}}

\def\txi{\tilde{\xi}}

\def\tga{\tilde{\gamma}}

\def\tl{\tilde{l}}

\begin{document}
\raggedbottom

\title{{\bf Unified description of cosmological and static solutions in
  affine generalized theories of gravity: vecton - scalaron duality and its applications.
 }}

\author{A.T.~Filippov \thanks{Alexandre.Filippov@jinr.ru} \\
{\small \it {$^+$ Joint Institute for Nuclear Research, Dubna, Moscow
 Region RU-141980} }}

\maketitle

\begin{abstract}

 We briefly describe the simplest class of affine theories of gravity in multidimensional space-times with symmetric connections and their reductions to two-dimensional \emph{dilaton - vecton gravity} field theories (DVG).  The distinctive feature of these theories is the presence of an absolutely neutral massive (or tachyonic) vector field (\emph{vecton}) with essentially nonlinear coupling to the dilaton gravity (DG).  We show that in DVG the vecton field can be consistently replaced by an effectively massive scalar field (\emph{scalaron}) with an unusual coupling to dilaton gravity. With this vecton - scalaron duality, one can use methods and results of the standard DG coupled to usual scalars (DGS) in more complex \emph{dilaton - scalaron gravity} theories (DSG) equivalent to DVG.  We present the DVG models derived by reductions of multidimensional affine theories and obtain one-dimensional dynamical systems simultaneously describing   cosmological and static states in any gauge.  Our approach is fully applicable to studying static and cosmological solutions in multidimensional theories as well as in general one-dimensional DGS models. We focus on global properties of the models, look for integrals and analyze the structure of the solution spaces. In integrable cases, it can be usefully visualized by drawing a `topological portrait' resembling phase portraits of dynamical systems and simply exposing  global properties of  static  and cosmological solutions,  including horizons, singularities, etc.  For analytic approximations we also propose an integral equation well suited for iterations.
\end{abstract}

\section{Introduction}
    The present observational data strongly suggest that Einstein's
    gravity must be modified. The combination of data on dark energy (DE) and  growing evidence for some sort of inflation have generated a wide spectrum  of such modifications
   (see, e.g., \cite{Starobin} - \cite{Martinec}).
   Superstring ideas suggested natural modifications but, in view of the serious mathematical problems of the present string theory, strictly definite predictions about concrete modifications of gravity are not yet available. Moreover, the phenomenon of dark energy was not predicted by string theory and its origin in the stringy framework  proved to be rather difficult to uncover and understand.
   The problem of dark energy in string theory looks very deep and is related to many other complex issues of (non-existing) quantum cosmology but it also led to some beautiful and exciting speculations, like eternal inflation and multiverse (see, e.g.,  \cite{Mavr} and \cite{Bousso}).  On the other hand,
   if we modestly try first to find a natural place for DE in classical cosmological models, which inevitably are essentially nonlinear and non-integrable, we better return to recalling the origin of general relativity and look for some options abandoned or not found
   by its creators.

   For this reason,  simpler modifications of gravity that affect only the gravitational sector are also popular, e.g., \cite{Nojiri} - \cite{Brax}. In essence, these modifications reduce to the standard Einstein gravity supplemented by some number of scalar bosons (the first example of such a modification was the old Jordan - Brans - Dicke theory). The main problem of this approach is that the origin of these scalar bosons is quite obscure,  and there is no theoretical principle regulating their coupling to gravity. Of course, there exist some phenomenological and theoretical constraints but in general the spectrum of these models is too wide.\footnote{Restricting consideration to homogenous cosmologies, one can find that in dimensionally reduced supergravity (superstring) theory there emerge many massless scalar bosons that couple to gravity only, see, e.g., \cite{Lidsey}, \cite{Copeland}. }
         The modification proposed and  studied in \cite{ATF} - \cite{ATFp}
    satisfy some principles of geometric nature based on Einstein's idea (1923)\footnote{References to his papers as well as to related work of Weyl and Eddington can be found  in \cite{ATF} - \cite{ATFg}. } to formulate the gravity theory
   in a non-Riemannian space with a symmetric connection determined by   a special variational principle involving a `geometric' Lagrangian. This Lagrangian is  assumed to be a function of the generalized Ricci curvature tensor as well as of other fundamental tensors and is varied in the connection coefficients. A new interpretation and  generalization of this approach
   was developed in \cite{ATFn} - \cite{ATFg} for an arbitrary
   space-time dimension $D$.

   The connection coefficients define symmetric and antisymmetric parts of the Ricci tensor ($s_{ij}$ and $a_{ij}$)  and a new vector $a_i$.
   Supposing that in the pure geometry there is no dimensional fundamental constants (except the velocity of light relating space to time) we choose geometric Lagrangians giving dimensionless geometric action. The geometric variational principle puts further bounds on the geometry and, in particular, relates $a_i$ to $a_{ij}$.  To define a metric tensor we have to introduce some  dimensional constant. Then we can find a physical Lagrangian depending on this dimensional constant and on some dimensionless parameters. Thus obtained  theory supplements
   the standard general relativity with dark energy
   (the  cosmological term, in the limit  $a_i = a_{ij} =0$), neutral massive (or tachyonic) vector field proportional to $a_i$ (vecton) and, after dimensional reductions to $D=4$,
   with $(D-4)$  massive (or tachyonic) scalar fields.

   Further properties of the theory depend on a concrete choice of
   the \emph{geometric Lagrangian}. The resulting
   physical theory can be described by the corresponding
   \emph{effective Lagrangian} that depends on the metric, vecton
   and scalar fields. In the geometric theory, there are no dimensional
   constants, while any fundamental tensor has dimension of some power
   of length (assuming $c=1$).
   We proposed a class of geometric Lagrangian
   densities depending on the symmetric  and antisymmetric parts of
   the generalized Ricci tensor, $s_{ij}\,$, $a_{ij}\,$,
   and on a fundamental vector obtained by contracting
    the connection coefficients $\Gamma_{ij}^k\,$, i.e., $a_i \sim \Gamma_{ik}^k\,$.
    By requiring the geometric `action' (the space-time integral of the geometric Lagrangian density)
   to be dimensionless we can enumerate all possible actions,
   see \cite{ATFn}.
     The corresponding  Lagrangian may be defined as
   the square root of an arbitrary linear combination of these densities
   or a linear combination of their square roots.

  The most natural density of this sort in any dimension is
  the square root of $\det(s_{ij} + \bl a_{ij})$, where $\bl$
 is a number.\footnote{Einstein used as the Lagrangian
 Eddington's scalar density $\sqrt{|\det r_{ij}|}\,$,
 where $r_{ij} \equiv s_{ij} + a_{ij}\,$.}.
     The effective physical Lagrangian is the sum
     of the standard Einstein term, the vecton mass  term,
 and the term proportional to $\det(g_{ij} + l f_{ij})$
 to the power $\nu \equiv 1/(D-2)$, where  $g_{ij}$ and  $f_{ij}$
 are the metric and the vecton field tensors \emph{conjugate}
 to $s_{ij}$ and $a_{ij}\,$,\footnote{This unusual construction
 introduced by A.~Einstein is described and generalized in
 \cite{ATF} - \cite{ATFg}, see also Appendix.}
 $l$ is a parameter related to  $\bl$.
 The last term has the dimensional multiplier,
which in the limit of small field $f_{ij}$ produces the cosmological
constant. For $D=4$ we therefore have the term first introduced by
Einstein but now usually called the Born - Infeld or brane Lagrangian.
For $D=3$ we have the Einstein - Proca theory, which is very
interesting for studies of nontrivial space topologies.

     The simplest dimensional reductions from $D>4$ to $D=4$ produce $(D-4)$ scalar fields and thus the complete theory is rather complex, even at the classical level.
   Its spherically symmetric sector is described by a much simpler
   (1+1)-dimensional dilaton gravity coupled to one massive vector
   and to several scalar fields. This dilaton gravity coupled
   to the vecton and massive scalars as well as its further
   reductions to one-dimensional `cosmological' and `static'
   theories were first formulated in \cite{ATF} - \cite{ATFg},
   and here we begin systematic studies of their
   general properties and solutions. These studies can be somewhat
   simplified by transforming the vecton field into a new massive scalar field, which is possible (on the mass shell) in the
   two-dimensional reductions.\footnote{
   In addition to some formal simplifications, the transformation
   may shed a new light on some theoretical problems, especially, in
   cosmology. Indeed, new cosmological theories operate with
   several exotic scalar fields (inflatons, phantoms, tachyons), which
   are usually introduced \emph{ad hoc}. In the models considered
   here, such `particles' appear as effective fields in a consistent theoretical framework.}
  As distinct from the normal scalar matter fields, the scalaron has
  a different coupling to gravity and may have abnormal signs of
  the kinetic term (phantom) or mass term (tachyon). Nevertheless,
  some general methods used in DGS models can be successfully applied
  to DG coupled to abnormal scalar fields. Thus, the main purposes of the present study are: to essentially modernize available tools for constructing various approximation schemes, to understand most important qualitative features of the global solutions, and to
  look for possible integrable approximation in simple one-dimensional reductions of the general theory.

 In this paper, we consider the simplest \emph{geometric Lagrangian},
  \be
 \label{1.1}
 {\Lag}_{\rm geom} = \sqrt{-\det(s_{ij} + \bl a_{ij})}\,
 \equiv \, \sqrt{-\Delta_s}  \,,
  \ee
 where the minus sign is taken because $\det(s_{ij})<0$
 (due to the local Lorentz invariance) and we
 naturally assume that the same is true for
 $\det(s_{ij} + \bl a_{ij})$
 (to reproduce Einstein's general relativity
       with the cosmological constant in the limit
 $\bl \rightarrow 0$).
 Following the steps of Ref.~\cite{ATFn} or using results described
 in Appendix~7.1 one can derive the
 corresponding \emph{physical Lagrangian}
  \be
 \label{1.2}
 \Lag_{\rm phys} =  \sqrt{-g} \,
 \biggl[ -2 \Lambda \,[\det(\delta_i^j +
 l f_i^j)]^{\nu} +  R(g) -
 m^2 \, g^{ij} a_i a_j \biggr] \,, \qquad
   \nu \equiv 1/(D-2) \,,
 \ee
  which should be varied with respect to the metric and the
  vector field; $m^2$ is a parameter depending on the
  chosen model for affine geometry and on $D$
  (see \cite{ATF} - \cite{ATFg}). This parameter can be
  positive or negative and we often use notation $m^2 \equiv \mu$.
       When the vecton field vanishes,
  we have the standard Einstein gravity
  with the cosmological constant. For dimensional
  reductions from $D \geq 5$ to $D=4$,
  we can obtain the Lagrangian describing
 the vecton $a_i$, $f_{ij} \sim \p_i a_j - \p_j a_i$
 and $(D-4)$ scalar fields $a_k, \, k = 4,..,D$.
  Note that, for $D=3$, the Lagrangian
 (\ref{1.2}) is bilinear in the vecton field, and in the
 approximation $a_i = 0$ it gives the three-dimensional gravity with the cosmological term. The three-dimensional gravity was
 studied by many authors (see, e.g., \cite{Carlip}, \cite{Witten1}
 and references therein)
 and this may significantly simplify the study of solutions of the
 new theory.

 We first consider the simplest spherical dimensional reduction
 from the $D$-dimensional theory (\ref{1.2}) to the two-dimensional
 dilaton-vecton gravity (DVG) and then further
          reduce it to one-dimensional static
 or cosmological theories. The next step is considering
          cylindrical reductions of the $D$-dimensional theory
 to more complex
          two-dimensional dilaton gravity theories coupled to
          several scalar fields.
 For simplicity, we consider
 this dimensional reduction only in
          the dimensions four and three.
 As is well known, in the $D=4$ case, in addition to the dilaton,
 there appear `geometric' $\sigma$-model scalar fields and vector
 pure gauge fields that produce an effective potential
 introduced in \cite{ATFr}.
 These two-dimensional scalars look like scalar `matter' fields
 obtained by dimensional reductions from higher dimensional supergravity
 theory. Like those fields, they are classically massless.
 However, in general, the $\sigma$-model coupling of the scalar fields
 is supplemented by the mentioned effective potential depending on
 the scalar fields and on
 the `charges' of the `geometric' pure gauge fields. This theory is
 rather
 complex and not integrable even after reduction to dimensions 1+0 or 0+1
 (see \cite{ATFr}). Adding the vecton coupling makes the theory
 even more complex and in this paper we consider only the simplest
 variants.

  Below, we mainly discuss low-dimensional theories
 ($D=1,2,3,4$) that allow us to consistently treat some solutions
 of realistic higher dimensional theories.
   We mostly concentrate on mathematical problems and do not
 pay much attention to physics meaning of the obtained solutions.
 Note also that
 in the context of modern ideas on inflation, multiverse, etc.,
 the main parameters of a fundamental theory of gravity cannot
 be theoretically determined. In particular, we do not know the sign
 and the magnitude of the cosmological constant and of other parameters
 giving a scale of length. Even the dimension and signature of the
 space-time should be considered as
 a free parameters that can be estimated in the context of a concrete
 scenario of the multiverse evolution or by anthropic considerations.
 In practice, this means that we should regard theories with any
 parameters in any space-time dimension as equally interesting,
 at least, theoretically.

 \section{Dimensional reductions of the generalized gravity theory}
 Let us outline the main reductions of the model (\ref{1.2}) in
 the dimensions $D=3, 4$.\footnote{A careful consideration
 of the general successive dimensional reductions of Einstein's gravity from  $D=4$ to $D=1$ is given in \cite{Chandra}, \cite{Gibbons}. For a review of the standard spherical and cylindrical reductions and solutions see \cite{Step}.}  Due to natural space and
 time restrictions we only give an overview of main points.
 First, consider a rather general Lagrangian in the
 $D$-dimensional \emph{spherically symmetric} case
 ($x^0 = t$, $x^1 = r$):
 \be
 ds_D^2 = ds_2^2 + ds_{D-2}^2 =
 g_{ij}\, dx^i\, dx^j \,+ \,
 {\f}^{2\nu} \, d\Omega_{D-2}^2 \, ,
  \label{2.1}
   \ee
  where $\nu \equiv (D-2)^{-1}$.
  The standard spherical reduction
  of (\ref{1.2}) gives the effective Lagrangian\footnote{
  We suppose that the fields $\f$, $a_i$ are dimensionless while
  $[t] = [r] = \textrm{L}$ and thus $[f_{ij}] = \textrm{L}^{-1}$,
  $[R] = [k_{\nu}] = [X] = [m^2] = \textrm{L}^{-2}$. For more details on our dimensions see Appendix.}
  the first three terms of which describe the standard spherically
  reduced  Einstein gravity:
  \be
 \label{2.2}
 \Lag^{(2)}_D =  \sqrt{-g} \,  \biggl[\f R(g) +
 k_\nu \, \f^{1-2\nu} + W(\f)\, (\nabla \f)^2 +
  X(\f, \textbf{f}^{\,2}) - m^2 \f \, \textbf{a}^2
 \biggr] \,.
 \ee
 Here $a_i(t,r)$ has only two non-vanishing components
 $a_0, a_1$, $f_{ij}$ has just one independent component
 $f_{01} = a_{0,1} - a_{1,0}$; the other notations are:
 $\textbf{a}^2 \equiv a_i a^i\,\equiv g^{ij} a_i a_j\,$,
 $\textbf{f}^{\,2} \equiv f_{ij} f^{ij}\,$,
 $k_\nu \equiv k(D-2)(D-3)$, $W(\f) = (1-\nu) /\f$
 and, finally, \footnote{
       The expression for the determinant in Eq.(\ref{1.2})
       in terms of $f_{ij}$ for $D=4$ written in
       \cite{ATFn} contains also the term of the fourth order in
       the vecton field. It is not difficult to see that
       in both spherical and cylindrical reductions (see
       the end of this Section)
       this term vanishes and thus Eq.(\ref{2.3})
   is valid. In fact, this is true in any dimension $D$.}
 \be
 \label{2.3}
 X(\f, \textbf{f}^{\,2}) \equiv
  -2 \Lambda \f \,\bigg[1 +
  \half \lambda^2 \textbf{f}^{\,2}\,\biggr]^{\,\nu} \,,
 \ee
    where, the parameter $\lambda$ is related to
    dimensionless parameter $l$ in (\ref{1.2}) but
    $[\la] = \textrm{L}$ .

 Sometimes, it is convenient to transform away the dilaton
 kinetic term by using the Weyl transformation, which in our
 case is the following ($w^{\pr}(\f) / w(\f) = W(\f)$):
 \be
 g_{ij} = \hat{g}_{ij} \, w^{-1}(\f) ,\,\,\,\,\,
 w(\f) = \f^{1-\nu}, \,\,\,\,\,\,
 \textbf{f}^{\,2} = w^2 \,\hat{\textbf{f}}^{\,2} \,,\,\,\,\,\,
 \textbf{a}^2 \,=\, w \, \hat{\textbf{a}\,}^2 \,.
 \label{2.4}
 \ee
 Applying this transformation to (\ref{2.2}) and omitting the
 hats we find the transformed Lagrangian
  \be
 \label{2.5}
 {\hat{\Lag}}^{(2)}_D  =  \sqrt{-g} \,  \biggl[\f R(g) +
 k_\nu \, \f^{-\nu} -
 2 \Lambda \f^{\nu} \,\bigg(1 +
  \half \lambda^2 \f^{2(1-\nu)}
  \textbf{f}^{\,2}\,\biggr)^{\,\nu}  -
  m^2 \f \, \textbf{a}^2 \biggr] \,.
 \ee
   When $D=3$ we have $\nu = 1$, $k_\nu =0$, Weyl's transformation
 is trivial and the Lagrangian is
  \be
 \Lag^{(2)}_{3}  =  \sqrt{-g} \, \f \biggl[ R(g) -
  2 \Lambda -  \lambda^2 \Lambda  \, \textbf{f}^{\,2} -
  m^2 \, \textbf{a}^2 \biggr] \,.
  \label{2.6}
 \ee

 These two-dimensional reductions are essentially simpler than
 their parent higher dimensional theories.
 In particular, we show that the massive vecton field theory
 can be transformed into a dilaton - scalaron gravity (DSG)
 model which is easier to analyze.
 Unfortunately, these DSG models and  their further
 reductions to dimension one (static and cosmological reductions)
 are also essentially non-integrable. It is well
 known that the massless case, being a pure dilaton gravity,
 is classically integrable (see, e.g., \cite{ATF1} and reference therein). Having this in mind,  we will attempt to find some additional integrals of motion
 in similar vecton models.

 The next simplified theory is obtained in a
 \emph{cylindrically symmetric} case. We consider here only
 $D=3$ and $D=4$. The general cylindrical reduction
 was discussed in detail in
 \cite{ATFr} and here we only summarize the main results.
 The most general cylindrical Lagrangian can be derived
 by applying the general Kaluza reduction to $D=4$. The
 corresponding metric may be written as
 \be
 ds_4^2 = (g_{ij} + \f \,\sigma_{mn} \,\f_i^m \f_j^n) \, dx^i dx^j +
  2 \f_{im} \, dx^i dy^m + \f \, \sigma_{mn} \, dy^m dy^n \, ,
 \label{2.7}
 \ee
 where $i,j = 0,1$, $m,n = 2,3$, all the
 metric coefficients depend only on the $x$-coordinates ($t,r$),
 and $y^m =(\f, z)$ are coordinates on the two-dimensional
 cylinder (torus). Note that $\f$ plays the role of the dilaton and
 $\sigma_{mn}$ ($\det \sigma_{mn} = 1$) is the so-called
 $\sigma$-field.
 The reduction of the Einstein part of the four-dimensional
 Lagrangian, $\sqrt{-g_4} \, R_4 $, can be written as:
 \be
 \cL^{(2)}_{4 {\rm c}}= \sqrt{-g} \, \biggl[ \f R(g) +
 {1 \over 2\f} (\nabla \f)^2  -
 {\f \over 4} {\rm tr} (\nabla \sigma \sigma^{-1}
 \nabla \sigma \sigma^{-1}) -
 {\f^2 \over 4} \sigma_{mn} \,\f^m_{ij} \,\f^{nij} \biggr] \, ,
  \label{2.8}
  \ee
 where $\f^m_{ij} \equiv \partial_i \f^m_j -
 \partial_j \f^n_i$.
 The Abelian gauge fields $\f_i^m$ are not propagating and their
 contribution is \emph{usually neglected}. In \cite{ATFr} we proposed
 to take them into
 account by solving their equations of motion and writing the
 corresponding effective potential (similarly to what we are
 doing below in the spherically symmetric vecton gravity).
 Introducing a convenient parametrization for $\sigma_{mn}$,
 \bdm
 \sigma_{22} = e^{\eta}\cosh\xi , \,\,\,\, \sigma_{33} =
 e^{-\eta} \cosh\xi, \,\,\,\, \sigma_{23} =
 \sigma_{32} = \sinh\xi \,,
  \edm
 one can exclude the gauge fields $\f_i^m$ and derive the
 effective action
 \be
 {\cal L}_{\rm eff}^{(2)} = \sqrt{-g} \, \biggl[ \f R(g) +
 {1 \over 2\f} (\nabla \f)^2  - {\f \over 2} [ (\nabla\xi)^2 +
 (\cosh \xi)^2  \,  (\nabla \eta)^2]  +
 V_{\rm eff}(\f, \xi , \eta) \biggr] \, .
 \label{2.10}
 \ee
 Here the first three terms are standard, while the effective geometric potential,
 \be
 V_{\rm eff} (\f, \xi , \eta) =
 -{\cosh \xi\over 2 \f^2} \biggl[Q_1^2 e^{-\eta}   -
 2 Q_1 Q_2 \tanh \xi + Q_2^2 e^{\eta}\biggr] \, ,
 \label{2.11}
 \ee
 depends on two new arbitrary real constants $Q_m$, which may be called
 `charges' of the Abelian geometric gauge fields $\f^m_{ij}\,$.
  This representation of the action is more convenient for writing
 the equations of motion, for further reductions to dimensions
 $(1+0)$, and $(0+1)$ as well as for analyzing special cases,
        such as $Q_1 Q_2 =0$, $\xi \eta \equiv 0$.\footnote{
        It is also closer to the
 original Einstein - Rosen equations for nonlinear
 gravitational waves, which
 can be obtained by putting $Q_1 = Q_2 =0$ and $\xi \equiv                         0$.
      When $Q_1 Q_2 \neq 0$, $\xi$ and $\eta$ cannot be
      identically zero.}

      The  static solutions  of the theory (\ref{2.11})
      with $Q_1 Q_2 \neq 0$ have horizons while the exact solutions
      discussed in  \cite{ATFr} for $Q_1 = Q_2 =0$ and
      non-vanishing $\sigma$-fields $\xi$, $\eta$
      have no horizons at all, in accordance with the general
       theorem of papers \cite{ATF1}.
      An interesting special case can be obtained if
      we choose $Q_1 \neq 0$, $Q_2 = 0$, $\xi \equiv 0$.
   Then, Lagrangian (\ref{2.10}) gives a standard dilaton gravity coupled to the scalar field $\eta$, with the potential depending
   both on the scalar field and dilaton. If $Q_1 \neq 0$,
   there exists a static solution with a horizon, which disappears
     when $Q_1$ vanishes. Of course, the horizon also exists
   in pure dilaton gravity,  when $\xi = \eta = 0$.

   In \cite{ATF1} we studied in some detail the models
   with the potentials independent of the scalar. In this paper, we
   demonstrate that interesting results can be derived in more
   general models with `separable' potentials
   $V(\f, \psi) = v_1(\f) v_2 (\psi)$.
   In particular, we show that one of the integrals of
   motion in the dilaton gravity coupled to massless scalars
   derived in \cite{ATF1} may exist also in some models with
   separable potentials that are of interest in the context
   of the present study.

  We see that the general cylindrical action is a very complex
  two-dimensional theory and even its one-dimensional reductions
  are rather complex and in general not integrable.
  Adding the vecton sector does not make it simpler and more
  tractable.
  Much more tractable is the \emph{three-dimensional} cylindrical
  space-time. The metric can be obtained by the obvious reduction
  of (\ref{2.7}),
  \be
 ds_3^2 = (g_{ij} +  \,\f_i\f_j) \, dx^i dx^j +
  2 \f_i \, dx^i dy + \f  \, dy^2 \, ,
 \label{2.12}
 \ee
 and the corresponding Einstein Lagrangian is simply
 \be
 \cL^{(2)}_{3{\rm c}} = \sqrt{-g \,\,\f} \,
 \{ R(g)  - {\f \over 4} \,\,\f_{ij} \,\, \f^{ij} \} \,,
 \qquad  \f_{ij} \equiv \f_{i,j} - \f_{j.i} \,.
 \label{2.13}
 \ee
 Using the equation of motion for $\f_{ij}$ and introducing
 the corresponding effective potential (see \cite{ATFr}
 and more general derivation below) we find the following
 two-dimensional dilaton gravity
 \be
 {\cal L}^{(2)}_{\rm eff} = \sqrt{-g} \, \{ \phi R(g) -
 8 Q^2 \, \phi^{-3} \}\, , \qquad  \phi \equiv \sqrt{\f} \,.
 \label{2.14}
 \ee
 As distinct from the cylindrical reduction of the
  four-dimensional pure Einstein theory corresponding
  to $ V_{\rm eff} = 0$ in (\ref{2.10}), this theory has a horizon and can easily be integrated.

 It is not difficult to include the vecton part into the cylindrically
 symmetric Lagrangians. In fact, the terms
 $-m^2 \f \, \textbf{a}^2 $,
 $X(\f, \textbf{f}^{\,2})$ in (\ref{2.3}) are invariant and
 have the same form in any dimension. Therefore we can simply add
 the expressions (\ref{2.10}), (\ref{2.13}) to the gravitational
 Lagrangians also in the cylindrical case. Note however that there exist different cylindrical reductions of the vecton
 potential $a_i$. For example, unlike the spherical case,
 these fields may be nonzero for $i = 0,1,2$ and correspondingly
        $e_1 \equiv f_{01} \equiv \p_0 a_1 - \p_1 a_0 \neq 0$,
        $e_2 \equiv f_{02} \equiv \p_0 a_2 \neq 0$,
        $h_3 \equiv f_{12} \equiv \p_1 a_2 \neq 0$.\footnote{
        Here $e_i \equiv a_{0i}, \, h_i \equiv \ve_{ijk} a_{jk}$.
        In the diagonal metric $g_{ij} = g_i \delta_{ij}$ the
        fourth-order term in the determinant from Eq.(\ref{1.2})
        is proportional to
        $(e_i h_i)^2 \exp{\sum 2 g_k}$ and is seen to vanish.
        This argument obviously works in any dimension.
        }
              We see that the component $a_2 \equiv  a_{\f}$
       of the vector field $a_i$ behaves like an additional scalar
       field and, in addition to the two-dimensional vector field
       $(a_0, a_1)$, we have up to three scalar matter fields,
       $a_2, \xi, \eta$, with a rather complex interaction to the
       dilaton gravity.
 This means that analyzing
  cylindrical solutions is more difficult than that of the
  spherical ones. Our consideration below is applicable to
  the simplest case of two potentials $a_0, \,a_1$ with one field $f_{01}\,$.

\section{Vector - scalar duality in two-dimensional vecton theory }
    In the dimension $D=2$ all fields
 (vector, spinor, ...) are practically equivalent to scalar
 ones. Such equivalence is widely known for massless Abelian
 gauge fields (see, e.g., \cite{ATF1} - \cite{VDATF1a}
 and references therein).
    The aim of
 this Section is to establish a standard map
 of massive Abelian vector fields to scalar fields.
 We do not attempt to present the most general results in this direction
 and therefore restrict  our consideration to a
 DVG coupled to scalar `matter' fields. This includes
 all the models of the previous sections.
  Suppose that in place of the standard Abelian gauge field
 term, $X(\f,\psi)\, \textbf{f}^{\,2}$, the Lagrangian contains
 a more general coupling of
 the gauge field $f_{ij} = \p_i a_j -\p_j a_i$
 to the dilaton and scalar fields, $X(\f, \psi; \textbf{f}^{\,2})$,
 where $\textbf{f}^{\,2} \equiv \,f_{ij} \,f^{ij}$.
 Using the Weyl transformation we write a fairly general two-dimensional Lagrangian in the form
  \be
  {\cal L}^{(2)} = \sqrt{-g}\, \biggl[ \f R + V(\f, \psi) +
  X(\f, \psi; \textbf{f}^{\,2}) +  Z_a(\f)\, \textbf{a}^2 +
  \sum Z (\f ,\psi)(\nabla \psi)^2  \biggr] \,,
    \label{3.1}
  \ee
  where the kinetic dilaton term $\sim \f^{-1} (\nabla \f)^2$
  is now absent
  while $V$ and $Z$ terms represent potentials
  and kinetic terms of scalar fields (like those in (\ref{2.10}),  (\ref{2.11})). We need not specify the number of
  the scalar matter fields and therefore
  omit the summation indices  for the fields $\psi$ and their
 $Z$-functions. For the fields, having positive kinetic
 energy, all the $Z$-functions
 in (\ref{3.1}) are negative and usually proportional to $\f\,$;
 we call them `normal matter' fields.

 In \cite{ATFHU}, we considered the massless case,
  $Z_a \equiv 0$, and proposed to use
 instead of (\ref{3.1})
 the effective Lagrangian not containing the Abelian gauge fields
 (for a detailed proof see \cite{VDATF1}, \cite{VDATF1a}).
 In this paper, we extend this approach to the massive vecton
 and show that the vector field can be effectively replaced
 by a scalar one (\emph{scalaron}) denoted by $q(t,r)$. This means that the
 equations of motion defined by Lagrangian (\ref{3.1}) can be
 obtained from the Lagrangian,
  \be
  {\cal L}_{\rm eff}^{(2)} = \sqrt{-g}\, \biggl[ \f R +
  V(\f, \psi) + X_{\rm eff}(\f, \psi, q) + Z_q(\f)(\nabla q)^2 +
  \sum Z(\f ,\psi)(\nabla \psi)^2  \biggr] \,,
  \label{3.2}
  \ee
  by applying a formally invertible transformation
  from the vecton fields $a_i$, $f_{ij}$ to the
  scalaron fields $\p_i q$, $q$.
    This can be constructed in a complete analogy to
    the massless case.

   First, consider the equations for the vecton fields
   derived from Lagrangian (\ref{3.1}):
  \be
 2 \p_j \,(\sqrt{-g}\,X^{\prime} f^{ij}) =
 Z_a(\f) \sqrt{-g} \, a^{i} \,,
 \qquad X^{\prime} \equiv X^{\prime}(\f, \psi;
 \textbf{f}^{\,2}) \,
 \equiv {{\p X} \over  {\p \, \textbf{f}^{\,2}}} \,,
  \label{3.3}
 \ee
  At this point, it is convenient to rewrite the Lagrangian and
  equations in the LC coordinates,
  \be
 ds^2 = -4\,h(u,v)\, du\,dv \,, \quad  \sqrt{-g} = 2h \,,
 \quad f_{uv} \equiv a_{u,v} - a_{v,u} \,,
 \quad  -2\textbf{f}^{\,2} = (f_{uv} / h)^2 \,.
  \label{3.4}
 \ee
 Before we pass to the LC Lagrangian we should derive
 the energy and momentum constraints using the general
 coordinates. Leaving only one extra field $\psi$
 we have (see, e.g. \cite{ATF},
 \cite{ATF1} - \cite{FM}):
  \be
 h \,\p_i(\p_i \f / h) \,=\,
  Z_a (\f) \, a_i ^2  +
  Z(\f ,\psi) \,(\p_i \psi)^2 , \qquad i=u,v.
 \label{4.2}
 \ee

 The other equations of motion can be derived from the LC
  transformed Lagrangian,
  \be
  \half {\cal L}^{(2)} = \f \, \p_u \p_v \ln|h| +
  h [V(\f, \psi) + X(\f, \psi, \textbf{f}^{\,2})] -
  Z_a (\f) \,a_u \,a_v -
  Z(\f, \psi) \, \p_u \psi \,\p_v \psi \,,
  \label{4.3}
  \ee
  simply by variations in  $h$,
 $\psi$, $\textbf{f}^{\,2}$,  $\f$. The most important
 equations are the following:
 \be
 \p_u \p_v \f \,+\, h \,\biggl[ X(\f, \psi, \textbf{f}^{\,2})\,
  +\, h {{\p X} \over {\p \, \textbf{f}^{\,2}}} \,
  {{\p \textbf{f}^{\,2}} \over {\p \, h}}
 \biggr] \, \equiv \,
 \p_u \p_v \f \,+\, h \,\biggr[ \, X -
 2 \textbf{f}^{\,2} {{\p X} \over {\p \,\textbf{f}^{\,2}}}\,\biggl] \,
 = \, 0 ,
 \label{4.4}
 \ee
 which allows us to define the effective potential
 $X_{\rm eff}(\f, \psi, q)$,
  and the vecton equations,
  \be
 \p_u (h^{-1} X^{\prime} f_{uv}) = - Z_a(\f) \, a_u \,, \qquad
 \p_v (h^{-1} X^{\prime} f_{uv}) =  Z_a(\f) \, a_v  \,,
 \label{3.5}
 \ee
 that are equivalent to (\ref{3.3}). Now we
  transform them into a standard scalar equation
  for a new  scalar field $q(u,v)$, which can be defined
 both in the LC and in general coordinates:
 \be
  q(u,v) \equiv  h^{-1} X^{\prime} \, f_{uv} \,, \qquad
 2 \sqrt{-g} \, f^{ij} \, X^{\prime} \equiv
 \ve ^{ij} \, q \,, \quad i,j = 0,1 \,.
\label{3.6}
 \ee
  It is not difficult to understand that so defined scalar $q$ is
  also invariant under the Weyl transformation  (\ref{2.4}).
  With this definition, we may regard $\textbf{f}^{\,2}$
  as a function of $q$,  which can be derived by solving (\ref{3.6}).
  Denoting  the solution by $z(q)$  and expressing $f_{uv} /h $ in terms of $\textbf{f}^{\,2}$ with the aid of (\ref{3.4}) we find an important relation:
  \be
 2 \, z(q) \,=\, -(q / \bar X^{\prime})^2 \,, \qquad
 \bar X^{\prime} \,\equiv \,{\p \over \p \,{ z(q)}}\,
 X [\f, \psi;  z(q)] \,.
 \label{3.7}
 \ee

 Returning to (\ref{4.4}), we see that the expression in the square brackets,
 with $\textbf{f}^{\,2}$ replaced by $z(q)$, is in fact our \emph{effective potential} $X_{\rm eff}(\f, \psi,  q)$. There exist several expression for it that allow one to derive the same equations of motion for the fields $h, \f, \psi, q$.
 In \cite{VDATF1}, \cite{VDATF1a} we obtained two
 equivalent expressions for $X_{\rm eff}$ by using
 (\ref{4.4}) and (\ref{3.7})):
 \be
 X_{\rm eff} (\f, \psi, q) \,=\,
  X[\f, \psi;  z(q)]
 -2  { z(q)}  \bX^{\prime}\, = \,
  X[\f, \psi;  z(q)] +
  q^2 /  \bX^{\prime} [\f, \psi;  z(q)] \,.
  \label{3.8}
 \ee
  From (\ref{3.7}) and (\ref{3.8}) we also derived
  the most compact and beautiful representation:
  \be
  X_{\rm eff}(\f, \psi, q) =
   X(\f, \psi; z) \,+\,
   q \,\sqrt{-2z}\,,
  \label{3.9}
 \ee
 where we intentionally write $z$ instead of $z(q)$.
 The reason is that
 \be
 \p_q \, X_{\rm eff} (\f, \psi, q) \,=\, (\bX^{\prime} -
 q / \sqrt{-2z}\,) \, \p_q \,z(q) \,+ \sqrt{-2z} \,=
 \sqrt{-2z} \,,
 \label{3.9a}
 \ee
 as follows from Eq.(\ref{3.7}).
 Now it is easy to write the equation for $q$.
 Using (\ref{3.5}), (\ref{3.6}) we find:
    \be
    a_u (u,v) = - Z_a^{-1} (\f) \, \p_u q (u,v) \,, \qquad
    a_v (u,v) = \, Z_a^{-1} (\f) \, \p_v q (u,v) \,,
   \label{3.5a}
    \ee
   \be
    \p_v (Z_a^{-1} \, \p_u q ) \,+\,
    \p_u (Z_a^{-1} \, \p_v q ) \,=\,
    -f_{uv} (q) = -h q /\bX^{\prime} \,=\,
    -h \p_q X_{\rm eff} (\f, \psi, q)
  \label{3.5b}
  \ee

  The last equation for $q(u,v)$  can be derived from Lagrangian (\ref{3.2}) by taking
  $Z_q = Z^{-1}_a$. Usually, $Z_a = -m^2 \f$ and then the scalaron
  kinetic term has the unusual form,
  \be
   Z_q \,(\nabla q)^2 = Z^{-1}_a \, (\nabla q)^2 =
   (\nabla q)^2 / (-m^2 \f) \,.
  \label{3.5c}
  \ee
  It is not difficult to derive all the equation of motion using
  effective Lagrangian (\ref{3.2}) with  $X_{\rm eff}$ and
  $Z_q$ given by  (\ref{3.9}), (\ref{3.5c}), and we list them here
  for further reference. In addition to the relations obtain above,
  one has to use the following easily checked identities:
  \bdm
 \p_h X_{\rm eff} (\f, \psi, q) \,=\, 0 \,, \quad
 \p_{\f} X_{\rm eff} \,=\,
 \p_{\f} \, X(\f, \psi, q) \,, \quad
 \p_{\psi} X_{\rm eff}  \,=\,
 \p_{\psi} \, X(\f, \psi, q) \,.
 \edm
  The first identity immediately follows from our definitions.
  In fact, it is a characteristic property of  $ X_{\rm eff}$
  and thus can serve to define it.
  Two other identities are derived like Eq.(\ref{3.9a}).

  Let us define the total effective potential,
  \be
  U(\f, \psi, q) \equiv V(\f, \psi) + X_{\rm eff} (\f, \psi, q)\,,
  \label{3.11}
  \ee
  and first write the LC version of the effective Lagrangian
  and of the constraints
 \be
  \half {\cal L}^{(2)}_{\rm eff} = \f \, \p_u \p_v \ln|h| +
  h U(\f, \psi, q) -
  Z_q (\f) \, \p_u q \,\p_v q  -
  Z(\f, \psi) \, \p_u \psi \,\p_v \psi \,,
  \label{3.12}
  \ee
   \be
 h \,\p_i(\p_i \f / h) \,-\,
  Z_q (\f) \,(\p_i q)^2 \,-\,
  Z(\f ,\psi) \,(\p_i \psi)^2 \,=\, 0. \qquad i=u,v.
 \label{3.12a}
 \ee
 The other equations are the same as used in
 the dilaton gravity coupled to scalars:\footnote{Below, we also
 use notation $\ln|h| \equiv F$, $h \equiv \epsilon \exp F$,
 where $\epsilon = +1$ for cosmological and $-1$ for
 static solutions.}
 \be
 \p_u \p_v \,\f + h \,U(\f, \psi, q) = 0 , \qquad
 \label{3.12b}
 \ee
  \be
 \p_u (Z_q \,\p_v q) + \p_v (Z_q \, \p_u q) +
 h \,\p_q U(\f, \psi, q) = 0 ,
 \label{3.12c}
 \ee
  \be
 \p_u (Z \,\p_v \psi) + \p_v (Z \,\p_u \psi) =
 \p_{\psi} [- h \, U + Z(\f, \psi) \,\p_u \psi \,\p_v \psi \,] ,
 \label{3.12d}
 \ee
 \be
  \p_u \p_v \ln|h| \,=\,
   \p_{\f} \,[- h \, U +
  Z_q (\f) \, \p_u q \,\p_v q  +
  Z(\f, \psi) \, \p_u \psi \,\p_v \psi \,] \,.
  \label{3.12e}
  \ee

  One of the last three equations is satisfied by solutions
  of the remaining five. The energy constraint, which is the sum
  the two constraints, plays a special role in the ADM
  Hamiltonian formulation \cite{ADM}. We do not discuss here these
  subtleties that will be clarified in considering
  the one-dimensional reductions of this
  two-dimensional  quasi-linear constrained system of the
  second order. The structure of this system allows one
  to find some integrable classes using
  simplifying assumptions about the potentials $U$ and $Z$.
  For some multi-exponential potentials $U$ and constant
  potentials $Z$, there exist integrable systems related
  to Liouville and Toda-Liouville ones (see \cite{ATF2},
   \cite{VDATF3}, \cite{VDATF2}). The pure Liouville case
   was completely solved but, for the Toda-Liouville, it is
   difficult to find an exact analytical solution of the
   two-dimensional constraints,
   even in the simplest $u_1 \oplus su_2$ case.
   This problem is easily solved in the one-dimensional
   (static or cosmological) case.

   For the spherical $D=3$ and $D=4$ Lagrangians
   in the Weyl frame (see (\ref{2.6}) and (\ref{2.5}), resp.),
   it is easy to derive the effective potential
   $\hU \equiv \, U / w(\f)$ and thus to find potential (\ref{3.11}):
    \be
     U(\f\,, q) =  \hU(\f,\, q) =
   -2\Lambda \f \,[\,1 + q^2 / 4\lambda^2 \Lambda^2 \,\f^2 \,],
   \qquad  D=3 \,,
      \label{3.14}
  \ee
  \be
   U(\f\,, q) =  \sqrt{\f} \, \hU(\f\,, q) = -2\Lambda \f \,
   [\, 1 + q^2 / \lambda^2 \Lambda^2 \f^2 \,]^{\half}
  \,+\, 2k\,, \qquad D=4 \,.
   \label{3.13}
  \ee
  We remind that the potential $Z_q$ is $D$-independent,
  $Z_q = -1/m^2 \f$, see Eq.(\ref{3.5c}).

   As was argued in \cite{ATF}, even the one-dimensional
   cosmological reduction of the pure scalaron theory
    with the potential (\ref{3.14})
   is not integrable. Thus we concentrate
   on searching for approximate potentials
    that allow us to find a wide enough class of analytic
   solutions to reconstruct exact ones by iterations.
    For instance, potential
   (\ref{3.13}) for large $q$
   is asymptotically very simple, in the standard frame we have
   $U = -2q/\lambda  + ...$.    As will be shown shortly,
   this asymptotic formula is valid in dimensions $D > 3$.
   On the other hand, the behavior of the $q$-dependent part of
   $U$ for small $q$ is given by (\ref{3.14}), if we multiply the
   $q^2$ term by $(D-2)$. This  universality allows us
   in what follows to regard  (\ref{3.14}), (\ref{3.13}) the generic scalaron potentials.

 \section{Cosmological and static reductions of scalaron theory}
 In this paper, we consider only the simplest reductions
 of the two-dimensional theory ignoring
 one-dimensional waves studied in our previous work
 \cite{ATFHU}-\cite{VDATF2}, \cite{VDATF3}. The simplest reduced gauge fixed equations can be directly
 derived by supposing that in the LC equations (\ref{3.12}) -
  (\ref{3.12e}) the fields $h, \f, q, \psi$ depend on one variable,
  which we denote $\tau \equiv (u+v)$. For the cosmological solutions
  this variable is identified with the time variable, $\tau = t$,
  while for the static states, including black holes (BH), it is
  the space variable, $\tau = r$. The only difference between the
  cosmological and static solutions is in the sign of the metric, $h_{\rm c} > 0$,  $h_{\rm s} < 0$.

  In our study of black holes and cosmologies we use
  the general diagonal  metric,
  \be
  ds_2^2 = e^{2\alpha (t, r)} dr^2  - e^{2\gamma (t, r)} dt^2 \,.
   \label{a.1}
 \ee
  Then, the static and cosmological reductions of our two-dimensional vecton theory (\ref{3.2}) can be presented by the Lagrangian
  (with the standard $Z(\f, \psi) = -\f$)
    \be
  \ep {\cal L}_{\rm v}^{(1)} = e^{\ep (\alpha - \gamma)} \f
  \biggl[\dpsi^2 - 2 \dal_{\ep} \, {\dvphi \over \f} -
  (1-\nu) \biggl({\dvphi \over \f}\biggr)^2 \biggr]
   - e^{\ep (\gamma-\alpha)} \mu \,\f \, a^2_{\ep} \,+\,
  \ep \, e^{\alpha + \gamma} \, \biggl[ V + X(\textbf{f}^{\,2})
  \biggr] \,.
  \label{a.2}
  \ee
   Here we omit a possible dependence of $V$ and $X$
   on $\f$ and $\psi$, denote $Z_a \equiv -\mu \f \equiv
   m^2$, and $\ep = \pm\,$.
   All the fields depend on $\tau = t$ ($\ep = +$)
   or on $\tau  = r$ ($\ep = -$). Finally,
   \bdm
    a_{+} = \,a_1\, (\tau)\,, \quad  a_{-} = \,a_0 \,(\tau) \,, \quad
   \alpha_{+} \equiv \alpha\,, \quad \alpha_{-} \equiv \gamma \,,
   \quad \dal_{\ep} = {d \over d\tau}\, \alpha_{\ep} \,, \quad
   \da = {d \over d\tau} \,a \,.
   \edm

   We see that the cosmological and static Lagrangians essentially
   coincide, the only difference being in the sign of the potential
   term and of the metric exponents as well. If $\ep = \pm $, the multiplier of the kinetic term,  $l_{\ep} \equiv e^{\alpha_\mp}$,
   is obviously a Lagrange multiplier varying of which produces the constraint equation, which is equivalent to vanishing of the Hamiltonian. In view of the  implicit dependence of  $\textbf{f}^{\,2}$ on $l_{\ep}\,$, it is much more convenient
   to first employ the canonical formulation and then identify a
   proper Lagrange multiplier. Alternatively, one can follow the steps made in previous Section but the canonical approach is simpler both technically and conceptually.

   Let us construct the effective Hamiltonian and Lagrangian of the
   vecton sub-Lagrangian,
     \be
     {\cal L}_a \equiv \,
     h \, X(\textbf{f}^{\,2})  \,-\,
     \tilde{\mu} \, \f \, a^2 \,;  \qquad \tilde{\mu} =
     e^{\ep (\gamma - \alpha)} \mu \,, \quad
     h = \ep \, e^{\alpha + \gamma} \,.
    \label{a.3}
    \ee
  Remembering our previous results we define the variable $y$
  and derive the momentum $p_a\,$,
   \be
     \half \lambda^2 \, \textbf{f}^{\,2}  \,= \,
     - \lambda^2 \biggl({\da \over h} \biggr)^2 = - y^2 \,,
      \qquad \da \equiv {1 \over \lambda} \, h\,y \,, \qquad
      p_a \equiv {\p {\cal L}_a \over \p{\da}} \,  =
      \lambda \, {\p X \over \p y} \,,
    \label{a.4}
    \ee
   where the last equation implicitly defines $y(p_a)$ . This allows us to write the partial Hamiltonian,
   \be
     {\cal H}_a(p_a\,, a) \equiv \,  p_a \,\da - {\cal L}_a =
     -h \, \biggl[ X - y  {\p X \over \p y} \biggr] \,+\,
     \tilde{\mu} \, \f \, a^2 \, \equiv -h X_{\rm eff} \,+\,
     \tilde{\mu} \, \f \, a^2 \,.
     \label{a.5}
    \ee
    Here $X_{\rm eff}$ essentially coincides with
    the effective potential derived in the previous section but now depends on $\f, \psi$ and $y(p_a)$. Applying the simplest possible canonical transformation,
    \be
    p_a \Rightarrow -2q\,, \quad  a \Rightarrow p/2 \,, \qquad
    X_{\rm eff}(p_a) \Rightarrow X_{\rm eff} (-2q)\,,
    \label{5a}
    \ee
    we see that the effective potentials (and the partial
    Lagrangians) obtained here and in Section~3 are identical.
    Indeed, applying the canonical transformation to Hamiltonian
    (\ref{a.5}), we find the partial Hamiltonian of the scalaron,
     ${\cal H}_s(p, q)$, and the corresponding Lagrangian,
    \be
       {\cal L}_s(p\,, q) \equiv \,  p \,\dq - {\cal H}_s (p,q) =
       h X_{\rm eff} \,+\, e^{\ep (\alpha - \gamma)}
       {\dq^2  \over \mu \f} \,,
       \label{a.6}
        \ee
   which can now be inserted into the complete Lagrangian instead of
   ${\cal L}_a\,$.

    Thus we have the complete DSG  Lagrangian, now denoting
    $l_{\ep} \equiv\, \exp(\alpha_{-\ep} - \alpha_{+\ep})$,
      \be
  \ep {\cal L}_q^{(1)} = l^{-1}_{\ep} \,
  \biggl[\f\, \dpsi^2 - 2 \dal_{\ep} \, \dvphi -
  (1-\nu) \, {\dvphi^2 \over \f} \,+\,
    {\dq^2  \over m^2 \f} \biggr]  +\,
   l_{\ep} \, \ep \,e^{2\alpha_{\ep}}\, U(\f, \psi, q) \,,
  \label{a.7}
  \ee
 where we also use notation (\ref{3.11}) and $\mu = m^2$.
  This form is more natural than
 (\ref{a.2}). First, the dependence on the Lagrangian multiplier
  $l_{\ep}\,$ is simple and standard,
 the kinetic part is quadratic in generalized velocities and can be
 made diagonal by a redefinition of the Lagrangian multiplier and
 velocities. In addition, we are free to make a convenient gauge choice and to choose the Weyl frame. For example, by making the shift
 $ \alpha_{\ep} \Rightarrow  \alpha_{\ep} -(1-\nu) \ln \sqrt{\f} $
  and redefining the potential by
 $U \Rightarrow \, \f^{\nu - 1} U$
 we remove the third term in (\ref{a.7}) and obtain the Lagrangian in
 Weyl's frame.\footnote{
 After this transformation, the theory defined by constrained Lagrangian (\ref{a.7}) can be compared to the
 one-dimensional reduction of the theory defined by
 equations (\ref{3.12}) - (\ref{3.12e}): we first derive the constraint,
 choose the LC metric by putting $l_{\ep}\,=1$, and then derive all
 the equations of motion. Note that the general one-dimensional
  Lagrangian (\ref{a.7}) gives more freedom in simplifying equations.
  For example, if there are only standard fields $\psi$, we may choose $l_{\ep} \equiv \tl_{\ep} \f$, $\txi \equiv \ln\f$ and then there
  will be no need to use Weyl's frame for simplifying the kinetic terms. }
   Then we can redefine $l_{\ep} \f \equiv \bl_{\ep}$,
  introduce the new field $\xi \equiv \,\f^2$ and
  finally rewrite (\ref{a.7}) in a simpler form,
  \be
  \ep {\cal L}_q^{(1)} = \bl^{-1}_{\ep}\,
  [\,\xi \dpsi^2 +\, m^{-2} \dq^2   - \dxi \, \dal_{\ep} \,] \,+\,
   \bl_{\ep} \, \ep \,e^{2\alpha_{\ep}}\, \xi^{\nu/2 -1} \,.
   U(\sqrt{\xi}\,, q, \psi) \,.
  \label{a.8}
  \ee
  Before applying it to studies of cosmologies
  and horizons in the scalaron theory we discuss the
  effective scalaron potential, corresponding to the $X$-potential
   (\ref{2.3}) in more detail.

  Using notation (\ref{a.4}) it is not difficult to find that
  $y(p_a)$ can be derived from the equation
  \be
   y = x\,(1-y^2)^{1-\nu} \,, \qquad
   x \equiv q / (-2\nu \lambda \Lambda \,\f) \,,
    \label{a.9}
  \ee
  where we choose the sign of $q$ so that $x>0$ and therefore $y>0$..
  This is a polynomial equation of order $(D-3)$ for $y^2$ if
  $D$ is even, and of order $2(D-3)$ for $y$ if $D$ is odd.
  This observation is not very useful if $D>4$, but some algebraic expressions for $y(x)$ can be written if $D=5, 6$. They are rather complex and difficult to use. However, it is not difficult
  to find good approximations for the solutions for small and
  large values of $x$:
  \be
  y = x \,[1- (1-\nu)\,x^2 + ...]\,; \qquad
  1 - y^2 = |x|^{-\sigma} - (\sigma/2)\,|x|^{-2\sigma} +... \,, \quad
  \sigma \equiv 1/(1-\nu) \,.
  \label{a.10}
  \ee
 Both expansions are applicable to all $\nu > 1$, the first is exact
 for $\nu = 1$ as  $y \equiv x$.

  To analyze the behavior of  $X_{\rm eff} (x)$ let us rewrite it
  in terms of the solution $y(x)$  of Eq.(\ref{a.9})
 \be
   X_{\rm eff} =  X - y  {\p X \over \p y} =
   -2\Lambda \f \,{x \over y} \biggl[1 - (1- 2\nu) y^2 \biggr]\,.
     \label{a.11}
  \ee
   This expression is convenient for deriving some exact solutions as well as for  discussing their general properties for arbitrary $D$. If $D=3$, then $y=x$, and we directly find the effective potential (\ref{3.14}). For $D=4$,  $x/y = \sqrt{1+x^2}$,   and thus the potential coincides with  (\ref{3.13}).
   An interesting feature of this effective potential
  is its invariance under
  rotations in the plane $(q, \lambda \Lambda \f)$.
  This invariance exists only in $D=4$. Unfortunately, the
  kinetic part does not respect this symmetry for any
  gauge choice.
    A different, simpler symmetry has the effective potential in $D=3$,
  which is equal to  (\ref{3.14})
  divided by $\f \equiv \sqrt{\xi}\,$.
  Then Lagrangian (\ref{a.8}) of the pure scalaron gravity,
  with $h \equiv \ep \,e^{2\alpha_{\ep}}\,$ and $\psi \equiv 0$,
  has a very simple form,
    \be
  \ep {\cal L}_q^{(1)} = \bl^{-1}_{\ep}\,
  [\,m^{-2} \dq^2   - \dxi \, \dal_{\ep} \,]  \,-\,
   \bl_{\ep} \,  2\Lambda h
   \,[\,1 + q^2 / 4\la^2 \La^2 \,\xi \,] \,,
  \label{a.12}
  \ee
  in which it is easy to find the scale symmetry $q \mapsto C\,q$,
  $\xi \mapsto C^{\,2} \,\xi$, $h \mapsto C^{\,2} h\,$.
  Such a symmetry itself does not allow to obtain and additional
  integral of motion. At most, it may signal of existence a special
  solution with a scaling-type dependence on $q, \xi, h$, but this is not our
  primary concern. Below, we will show that, in fact, the theory
  (\ref{a.12}) has an additional integral, but first we should complete
  our discussion of the general properties of the theories (\ref{a.8}).

  Applying asymptotic approximations (\ref{a.10}) to (\ref{a.11}) we prove
  the universality mentioned above:
  \be
   X_{\rm eff} =
      -2\Lambda \f \,[1 + q^2/4 \nu \la^2 \La^2 \f^2 \ +
      O(x^4)]\,,      \qquad
   X_{\rm eff} = - 2 \sqrt{q^2/\la^2}   + O(|x|^{-\sigma}) \,.
     \label{a.13}
  \ee
  The small $q$ approximations is exact for $\nu =1$,
  the large $q$ approximation is independent both of $\f$ and $\nu >1$,
  and the general structure of $ X_{\rm eff}$ is defined by
  a positive regular function, $v_{\nu} (x)$,
  \be
   X_{\rm eff} /(-2\Lambda \f) \, \equiv v_{\nu} (x) \,; \qquad
      v_{\nu}(0)=1 \,; \quad v_{\nu} (x) \simeq 2\nu x \,,
   \quad x\rightarrow \infty\,.
     \label{a.14}
  \ee
  Thus the effective potential in Lagrangian (\ref{a.8}) is
  \be
  U_{e} (\f, x) \equiv \, \xi^{\nu/2 -1} \, U(\sqrt{\xi}\,, q, \psi)
  = \,\f^{\nu -2} U =\, - 2 \La \f^{\nu -1}\, v_{\nu}\, (x) \,+\,
  k_{\nu}\, \f^{-(1 + \nu)} \,,
  \label{a.15}
  \ee
  and the exact expressions for $\nu = 1,\, 1/2$ are given in
  (\ref{3.14}), (\ref{3.13}).

   Let us discuss properties of  $v_{\nu}\,(x)$,
  which can be considered as the function of $q^2/\f^2
  \equiv q^2/\xi\,$ (it is expressed in terms of $y(x)$ by
  Eq.(\ref{a.11})). Using (\ref{a.9}) and (\ref{a.10})
  it is easy to find that
  \bdm
  {dy \over dx} =  {y \over x} \,
  {1 - y^2 \over 1 + (1 - 2\nu)\,y^2}  > 0 \,, \qquad
  y^{\pr} (0) = 1 \,, \quad  y^{\prime} (\infty) = 0 \,,
  \edm
    \be
   v_{\nu}^{\pr}(x) \,=\, 2\nu \, y(x) \,; \qquad
   v_{\nu}^{\pr}(x)> 0\,, \quad  v_{\nu}^{\pr \pr}(x) > 0\,.
    \label{a.16}
  \ee
  Thus $v_{\nu}(x)$ is monotonic concave function having simple
  expansions for $x \ll 1$, $x \gg 1$:
  \be
   v_{\nu}(x) \,=\, 1 + \nu x^2 + O(x^4)\,; \qquad
   v_{\nu}(x) \,=\, 2\nu x \biggl[1 +
   {1-\nu \over 2\nu}  x^{-\sigma} + O(x^{-2\sigma}) \biggr] \,.
    \label{a.17}
  \ee
   With such a simple and regular potential $U_e\,$,
   one might expect that at least
  qualitative behavior of the solutions of the theory (\ref{a.8})
  could be analyzed for small and large values of $x$.
  This is true if the theory is integrable. But it is not
  integrable, even in the simplest $D=3$ case when $v_1(x) = 1 + x^2$,
  $k_1 = 0$ and thus $U_e = -2\La (1+x^2)$ is linear in
  $q^2/\xi$. The integrability of this and of similar but more general theories is discussed in next Section and in a separate paper. Here we consider the scalaron
   canonical equations and find one additional integral of motion.

  Consider the general pure scalaron theory in which
  we neglect
  the curvature term $k_{\nu}\, \f^{-(1 + \nu)}$:
    \be
  {\cal L}_q^{(1)} = \,\bl^{-1}_{\ep}\,
  [\, \mu^{-1} \dq^2   - \dxi \, \dal_{\ep} \,]  \,-\,
   \bl_{\ep} \,  2\Lambda h\, \f^{\nu -1}\, v_{\nu}\, (x)\,,
  \qquad  \mu \equiv m^2 \,.
  \label{a.18}
  \ee
  We did not yet fix the gauge, $\bl_{\ep}$ is arbitrary.
  Varying in it we get the vanishing Hamiltonian,\footnote{
  Alternatively, we can use the standard Legendre transformation
  and write the Hamiltonian theory in the Lagrangian form,
  ${\cal L}(q, \dq) = p\,\dq - l H(p, q)$, which is most convenient
  for systems with constraints (see, e.g., \cite{Henn}).}
  \be
  {\cal H}_q^{(1)} \equiv \, \bl_{\ep}\, H_q^{(1)}
    = \bl_{\ep}\,[\,\mu p_q^2/4 - p_{\alpha} p_{\xi}  \,+\,
     2\Lambda h\, \f^{\nu -1}\, v_{\nu}\, (x)\,] \,=\, 0 \,,
  \label{a.19}
  \ee
  where we used the definition of the canonical momenta
  (here and below, $\al_\ep$ is replaced by $\al$),
  \be
   \dq  = \,\bl_{\ep}\, \mu p_q/\,2 \,, \quad
  \dxi = - \,\bl_{\ep} \,p_{\alpha}  \,, \quad
   \dal = - \,\bl_{\ep} \,p_{\xi} \,.
  \label{a.20}
  \ee
    We may use the parametrization
  invariance of the theory to choose a most convenient
  evolution parameter $\tau$, for example,
  $d\btau \equiv \,\bl_{\ep} \,d\tau = \dxi \, d\tau = d\xi$.\footnote{
  This choice is locally possible and is good as far as $\dxi \neq 0$.
  Here we ignore such subtleties.}
  Such a parametrization is equivalent to gauge fixing plus a coordinate
  transformation. The standard LC gauge choice in our present notation is
  $\,l_{\ep} = 1$, we mostly use the gauge
  $l_{\ep} \f \equiv \,\bl_{\ep} =1$ and keep the dot notation for
  $d /d\tau$. To further simplify the equations,
  consider the $\nu =1$ case and denote
  $\,2\Lambda h\, v_1\, (x) \equiv h \om \,(q^2/\xi)$. With
  this notation and with $\bl_{\ep}=1\,$, the Hamiltonian system is given by  equations (\ref{a.20}) plus
 \be
  \dop_q  =\,-2h \om^{\pr}(q^2/\xi)\, q/\xi \,,
  \quad
  \dop_{\xi} = h \om^{\pr}\,(q^2/\xi)\, q^2/\xi^2 \,, \quad
   \dop_{\al} =  -2h \om\,.
  \label{a.21}
  \ee

    It is easy to see that this Hamiltonian system has two integrals
  of motion. The first is the Hamiltonian (\ref{a.19}). The second
  one follows from the obvious relation
  $q \dop_q +2 \xi \dop_{\xi} = 0$:
    \be
  {d / d\tau} \,(q \,p_q + 2 \,\xi\, p_{\xi}) =
  q\, \dop_q + 2\,\dxi\, p_{\xi} = -2 h \om = \,\dop_{\al} \,,
  \label{a.22}
  \ee
  where we used  (\ref{a.20}), Hamiltonian constraint (\ref{a.19}),
  and the last equation in (\ref{a.21}). Integrating this relation
  gives the desired second integral. Thus we have two integrals:
    \be
  \mu p_q^2 - 4 p_{\alpha} \,p_{\xi}  + 4 h \om = 0\,; \qquad
  q \, p_q +2\, \xi\, p_{\xi} - p_{\al} = \,c_0\,.
  \label{a.23}
  \ee
  Knowledge of these integrals does not allow us to integrate our
  system but it is helpful in analyzing its properties.
  One may hope that, for some  particular potentials $\om$,
  a third integral can be found. For these reasons we call such systems \emph{partially integrable}.
      The present observation suggests a more general approach to
  constructing system with additional integrals presented in next Section.

 \section{Integrals and integrability in simple cases}
 Here we consider a general DGS with one scalar $\psi$
 that may be a standard  field or the scalaron:

 \be
 {\cal L}_{\rm dgs}^{(2)} = \sqrt{-g}\, \biggl[ \f R
 + Z(\f)(\nabla \psi)^2
 + V(\f,\psi) \biggr] \,.
  \label{a.24}
  \ee
 For the scalaron we have $\psi = q$, $Z = Z_q = -1 /(m^2 \f)$ and the potentials are given above. For the standard scalar $Z_{\psi} \sim -\f$, but some of the results presented below are applicable to more general $Z$-functions. In our notation,
 negative signs of $Z$ correspond to positive kinetic energies of
 the scalar fields but our classical consideration is fully applicable
 to both signs. The general model (\ref{a.24}) with a general
 potential $V$  is not integrable in any sense. One of the strong obstructions to integrability is the dependence of  $Z$ on $\f$, and the  usual simplifying assumption is that the $Z$-functions are independent of $\f$. With this restriction, there exists a class of `multi - exponential' potentials, for which the DGS theories with   any number
  of scalar fields can be reduced to the Toda - Liouville
  systems  and exactly solved.\footnote{
  This class includes all previously considered integrable
  two-dimensional DGS, which are reviewed in \cite{ATF2} (see
  also \cite{Kummer}).  The first DGS of the Liouville type (`bi Liouville'), which generalizes
  the so-called Jackiw \cite{Jackiw} and CGHS \cite{CGHS} models, was proposed and solved in papers \cite{ATF1}, the results of which are essentially generalized here.}
  For their  static - cosmological reductions,
    analytic solutions were explicitly derived. Here we try to expand this class of the models.

  First, consider the case of the massless vecton.
    In the pure scalaron DG we have, in the massless limit,
  $h \,\p_i(\p_i \f / h) = 0$, from which we find $h(u,v) = h(\tau)$,
  $\f(u,v) = \f(\tau)$, where $\tau = a(u) + b(v)$. This means that the
  theory is automatically reduced to dimension one and we can use
  the Hamiltonian formulation of Section~4 (in the LC gauge).
  It is clear $q$ is a constant, $q \equiv q_0$,
  as can also be seen from (\ref{3.5}), (\ref{3.6}). Now, using the above definitions
  of the momenta $p_{\alpha}, \,p_{\xi}\,$, constraint
  $p_{\al} \,p_{\xi} = h\om(\xi)$, and equation $\dop_{\al} = -2h \om(\xi)$,
  all of which are valid for arbitrary
  potential $\om(\xi)$, one can explicitly solve the
  equations of motion:
  \be
  \dop_{\al} = -2h \om\, = -2 p_{\al} \,p_{\xi}  \quad \Rightarrow
  \quad 2\dal p_{\al} = \dop_{\al} \quad \Rightarrow
  -\quad \dxi \equiv  p_{\al} = - c_ 1 h \quad \Rightarrow
  \quad c_1 \dh = - 2h \om \,,
  \label{a.25}
  \ee
 where $c_1$ is the integration constant. Then it follows the
 well known expression for $h$,
  \be
   c_1^2\,{dh \over d\xi} \,= \,2 \,\om(\xi)\,, \qquad
  c_1^2 \, h = 2\int d\xi \, \om(\xi) = 2\int d\f \,V(\f) \,
  \equiv N(\f) - N(\f_0)\,,
  \label{a.26}
  \ee
  if we recall that the transformation of $\xi$ to $\f$ induces the transformation of the potential, so that
   $d\xi \, \om(\xi) = d\f V(\f)$.
  Thus we find the explicit general solution of the general dilaton
  gravity, if we in addition solve the equation
  $\dxi = c_ 1 h(\xi)$ defining the $\tau$-dependence.
  Now, in our special case of the dilaton - scalaron gravity corresponding to the massless vecton, the only trace of the vecton is
  the dependence of the potential on $q_0$, the `charge' of
  the massless vector field, and on other parameters defining the
  theory. The general solution depends on the free independent parameters,
  ($c_1$, $\f_0$, $\tau_0$),\footnote{The physical parameters are
  defined by the equation for horizons,  $N(\f) - N(\f_0) = 0$.
  When this equation has a unique non-degenerate solution, as in the case
  of the Schwarzschild black hole, $N(\f_0)$ can be related to its
  `mass';  the case of many solutions, $\f_n$, the physical interpretation is much reacher, see, e.g. \cite{Chandra}, \cite{Frolov}. }
  which can be hidden by rescaling
  of $h$ and $\f$ and by shifting $\tau$. Then the `portrait' of the solution is a curve
  in the $(h, \f)$ plane significantly depending on  $q_0$.
        The most important feature of this portrait
  is the structure of the set of the horizons. For us, the most interesting
  is the dependence of this portraits on the parameters of the parent
  DSG: $q_0, \Lambda, \lambda, \nu$. Characteristic properties
  of the portraits are: the number and structure of the horizons
  (simple, degenerate, singular) and the behavior of the solutions near horizons and singularities. A more interesting portrait of an integrable DGS is described in Appendix.

   For DGS (\ref{a.24}), we can derive some convergent expansions of exact solutions near the horizons by applying the general approach proposed in \cite{FM}. Here we formulate its most compact form, which is convenient to use in general DGS with arbitrary $Z$, $V$
   ($V_{\psi} \neq 0$), and any number of scalars.
   For the sake of generality, we do not explicitly use the canonical formalism and thus somewhat change
   notation to the one resembling notation of \cite{ATF1}.
   We will also use the LC gauge and Weyl's frame having in mind that
   Eqs.(\ref{a.7})-(\ref{a.8}) show how to return to the standard frame
   and general gauge. Thus the system of one-dimensional equations
   can be obtained by dimensional reduction of
   (\ref{3.12a})-(\ref{3.12e}) if we suppose that all unknown functions depend on one variable $\tau = u + v$. Denoting
   $F \equiv \ln h$ and introducing the new momentum-like variables
   $\chi, \eta, \rho$,
   \be
  \qquad    \quad \dot{\varphi} = \chi \,, \quad
 Z(\f)\, \dpsi = \eta \,, \quad Z(\f) \,\dF = \, \rho \,,
 \label{a.27}
 \ee
 we rewrite the main dynamical equations  and the constraint as
 \be
   \dchi +  hV  = 0\,, \quad
  2 \deta + hV_{\psi} = 0 \,, \quad
  \dot{\rho} +h (ZV)_{\f} = 0\,, \qquad
  \chi \rho + h ZV + \eta^2 = 0\,,
  \label{a.28}
  \ee
  where the lower indices $\f,$ and $\psi$ denote the corresponding partial derivatives. It is well known that the second or the third equation can be omitted\footnote{The full system (\ref{a.27})-(\ref{a.28}) is evidently over-complete and  may be called the extended system. It is useful when we look for integrals of motion. Its subsystems can be used for constructing some special solutions.}
  and actually we have two independent second order  differential equations and one constraint which is the first order equation. Therefore,  to completely solve this system we must search for two additional constraints canonically commuting with the Hamiltonian and with each other.\footnote{
  As far as we are interested in the classical theory we usually will look for integrability in the Liouville sense.}

  Now, recalling the previous section we change the variable $\tau$ to
  $\xi$ that is defined by the relations: $\chi d\tau = d\f \equiv Z d\xi$. Then we rewrite equations
  (\ref{a.27})-(\ref{a.28}) denoting the derivative $d/d\xi$ by the prime  and introducing useful notation:
   $U(\xi, \psi) \equiv Z(\f) V(\f, \psi)\,$,
   $H(\xi) \equiv \, h/\chi\,$, and $G(\xi) \equiv \,\eta/\chi \,$.
   The main independent equations for $\chi$, $\eta$, $\psi$, $H$ now have a very compact form:
  \be
  \prpsi = G \,, \quad \prH = -G^2 H\,, \qquad
  \prchi +\, U H = 0\,, \quad  2\preta +\, U_{\psi} H = 0\,.
    \label{a.29}
  \ee
  The extended system contains two equations for $\rho$
  (see (\ref{a.27})-(\ref{a.28})):
    \be
  \chi \prF - \rho = 0\,, \quad
  \prho + U_{\xi} H = 0 \,, \qquad
  \chi \prG = UH(G - U_{\psi} /2U)\,,
  \label{a.30}
  \ee
  where we add the explicit equation for $G$,
which may be used instead of the last equation in (\ref{a.29}).

    The system (\ref{a.29}) is most convenient for deriving the solutions near horizons and in asymptotic regions and for studying their general properties.
    For example, a very important property  of its solution is that
   $(\ln H)^{\pr} = - G^2 <0$.  This property does not depend on the potential and is true for any number of scalar fields provided that their  $Z$-functions are negative,  as was first shown in \cite{ATFp}.
   Indeed, in this case the constraint equation can be written as:
   \be
   \prPhi \equiv \,(\ln H)^{\pr} =
   - Z_0 \sum_{n=0}^N Z^{-1}_n (\xi)\, (\eta_n /\chi)^2   \,, \qquad
   \eta_n \equiv Z_n \, \dpsi \equiv \chi \, \prpsi \,.
  \label{a.30a}
  \ee
   For normal fields $Z_n < 0$ for all $n$, for some anomalous fields,
   like scalaron corresponding to tachyonic vecton,
    $Z$ may be positive and then the sign of $\prPhi$  depends on the concrete solutions. In case of the same signs, Eq.(\ref{a.30}) resembles the second law of thermodynamics and defines an `arrow of time' for our dynamical system. If this is not true, the theorem is violated in a very specific way. It may be an interesting point for  cosmological modeling.

   Our system of equations (\ref{a.29}) has other interesting properties. The general solution of the first two equations can be written in terms of integrals of the function $G(\xi)$:
     \be
  \psi (\xi) \,=\,  \psi_0 \,+\, \int_{\xi_0}^\xi G \,,   \qquad
  H(\xi) \,=\, H_0 \exp \biggl(-\int_{\xi_0}^\xi G^{\,2} \biggr) \,.
    \label{a.31}
  \ee
   Then, inserting these `solutions' into the third and the forth equations and integrating them we can write one integral equation for  $G(\xi)$ instead of  system (\ref{a.29}):
      \be
  G (\xi) \equiv {\eta \over \chi} \,=\,  \biggl(\,\eta_0 \,-\,
  \half \, \int_{\xi_0}^\xi U_{\psi} H \biggr) \,
   \biggl( \,\chi_0  \,-\,   \int_{\xi_0}^\xi U H \biggr)^{-1} \,,
    \label{a.32}
  \ee
   where $\psi(\xi)$ and $H(\xi)$ are given by Eq.(\ref{a.31}). As was discussed in \cite{ATF1} and \cite{FM}, the standard
   (regular and non-degenerate) horizon appears when
   $\chi_0 = \,\eta_0 = 0$. Then $h(\xi_0) = 0$ while
   $G(\xi_0) = U_{\psi} (\xi_0)/\,2U (\xi_0)$ is finite if
   $U_{\psi} \neq 0$. It follows that $G$, $\psi$, $H$ are finite and can be expanded in convergent series around $\xi_0$ if the potential
   $U(\xi, \psi)$ is analytic in a neighborhood of $(\xi_0, \psi_0)$.\footnote{
   This is shown in \cite{FM}. In \cite{ATFp} one can find a detailed discussion of regular solution with horizons, including a generalization of the Szekeres - Kruskal coordinates as well as examples of singular and degenerate horizons.}
   When $U_{\psi} \equiv 0$ there is the obvious integral of motion
   $\eta = \eta_0$. As can be seen from the above equations and was proved in  \cite{ATF1}, there is no horizon for $\eta_0 \neq0$
   but, in the case of $\eta_0 = 0$, we have $G \equiv 0$ and return to
   pure dilaton gravity with the solution (\ref{a.26}) always having horizons.

   In simple cases, the integral equation can easily be reduced to a differential one. For example, if $U = u(\xi)\, v(\psi)$ and $U_{\psi} = 2g \,U$, the integral equation can be reduced to the second-order
   differential equation, which is not integrable for arbitrary $u(\xi)$
   but is explicitly integrable if $U_\xi = g_1 U$. This result is quite natural as in this case there exist two additional integrals,
   $\eta = g \chi + \eta_0$ and $\rho = g_1 \chi + \rho_0$ and therefore
   the most direct approach is to use the extended differential system.
   In non-integrable cases or when there is only one additional integral, the integral equation still can be a quite useful analytical tool, which we describe  in a separate publication. Here, we only briefly outline a generalization of the approach of Ref.\cite{ATF1} to finding potentials $U(\xi, \psi)$ for which the extended differential system has additional integrals.

  Generalizing the approach of \cite{ATF1} and the above remarks
  about possible integrals of motion let us collect those equations which can generate such integrals:
   \be
  0 \,=\, \rho \,+\, U H \,+\, \eta^2 /\chi \,= \,
   \prchi +\, U H \,=\,  \preta +\, U_{\psi} \,H/2 \,=\,
    \prho + U_{\xi} H \,= \,0\,,
  \label{a.33}
  \ee
  \be
  0 \,=\, \psi \,\preta +\, \psi \,U_{\psi}\, H/2 \,=\,
  \eta \,\prpsi - \eta^2 /\chi \,= \,
   \xi \,\prho + \xi\, U_{\xi}\, H \,= \,
   \xi^{\pr} \rho \,-\, \rho \,= \,0 \,,
    \label{a.33a}
  \ee
  where the first equation in (\ref{a.33}) is the energy constraint, which we denote  $E_0$ and the next items in this chain of equations
  are denoted by $E_i\,$, $i = 1,...,7$. Now, taking the sum
  $\sum_0^7 c_i E_i$ with $c_4 = c_5 =c_6 =c_7 = c_0$ we find that
  the solutions of equations (\ref{a.33}) satisfy the identity
   \be
   [c_1 \chi +\, c_2 \,\eta +\, c_3 \,\rho +\, c_4 (\psi \,\eta +\, \xi \,\rho)]^{\pr} =
   -H [(c_1 +\, c_4)\,U +\, c_2 \,U_{\psi}/2 +\, c_3 \,U_{\xi} +\,
   c_4 (\psi \,U_{\psi}/2 + \xi \,U_{\xi})]\,.
  \label{a.34}
  \ee
  Therefore, if the r.h.s. identically vanishes, the l.h.s generates the integral of motion,
   \be
   c_1 \chi +\, c_2 \,\eta +\, c_3 \,\rho +\, c_4 (\psi \,\eta +\, \xi \,\rho) = I_1 \,.
   \label{a.35}
   \ee
    This means that for the potentials $U(\xi, \psi)$, satisfying the partial differential equation
   \be
   (c_1 +\, c_4)\,U +\, c_2 \,U_{\psi}/2 +\, c_3 \,U_{\xi} +\,
   c_4 (\psi \,U_{\psi}/2 + \xi \,U_{\xi})\,=\,0 \,,
  \label{a.36}
  \ee
  there exist the corresponding integral of equations (\ref{a.33}). All the  above
  integrals can be obtained by applying this theorem.\footnote{
  An interesting exception is the integral (26) in Ref.\cite{ATF1}
  which apparently requires a more general method for its explanation.
    E.A.Davydov formulated a group theoretical approach for deriving  integrals of motion in general DGS theories, which apparently allows to obtain more general integrals than those discussed here.}
    The solution of Eq.(\ref{a.36}) depends on an arbitrary function of one variable. Using this fact  it is possible, in some simple cases, to derive one more integral.

   In the above example of the potential $U = U_0 \exp(2g\psi + g_1\xi)$
   with two additional integrals we immediately find the equation for $\chi$,
   \be
   \prchi = (g^2 + \,g_1)\, \chi \,+\, (\rho_0 + 2\, \eta_0 \,g)\, +\,
   \eta_0^2/\chi \,,
   \label{a.37}
   \ee
   by solving of which we explicitly express $\xi$, $\eta$, $\rho$, $h$ and $\psi$ as functions of $\chi$. This is sufficient for finding the portrait of this physically interesting system, which will be presented elsewhere.

  To demonstrate the problems, which remain even for apparently simpler systems, we consider the potential $U(\psi)$ also having two additional integrals. The obvious linear integral is $\rho = \rho_0$. To obtain one more integral suppose that $\psi U_{\psi} = 2 g U$. Then we have the additional bilinear integral of the three differential equations:
    \be
    \psi \eta -(g +1)\,\chi + \rho_0 \,\xi = I_0\,, \qquad
    \prchi = \rho_0 + \eta^2/\chi\,, \quad
    \eta \,\prpsi = \eta^2 /\chi  \,, \quad \psi \,\preta = g \prchi \,.
   \label{a.38}
   \ee
   We can exclude $\psi$ (or, $\eta$) and thus get two equation for $\chi$ and $\eta$  (or, $\psi$).  However, this is not an integrable dynamical system for two functions because of its explicit dependence on $\xi$. If $\rho_0 = 0$ in the expression for the integral $I_0$, it can be explicitly integrated, like the previous case, but this gives
   only a `partial' solution. This is a typical problem -- bilinear integrals having $\xi$-depending terms that describe a sort of a `back-reaction' of gravity on matter.

    A detailed comparative analysis of most interesting integrable and partially integrable DGS systems will be given is a separate publication.
    In conclusion of this Section we  summarize its main points.
    Dilaton gravity with scalars is in general not integrable even with formally sufficient number of integrals of motion. The models with massless scalars qualitatively differ from the scalaron models (DSG) that inevitably include non-integrability. Fortunately, in some physically important cases the non-integrable systems are partially integrable and therefore can be effectively studied, at least qualitatively. The solutions near horizons and singularities can be
    studied analytically, using exact expansions as well as iterations of the master integral equation. On the other hand, our approach to constructing systems with additional integrals may help to find integrable  or partially integrable systems qualitatively close to the realistic ones.

 \section{Summary and outlook}
 The main new results of this paper are the following.
 In Section~2, the standard spherical reduction of
 the $D$-dimensional DVG and its Kaluza-like cylindrical reduction
 in the dimensions 3 and 4 are briefly summarized.
 In Section~3 we derive the transformation of the two-dimensional DVG into the equivalent DSG. This opens a way for applying
 to the vecton theory some methods
 developed in two-dimensional dilaton gravity coupled to
 scalar fields. In particular, the
 nonlinear kinetic terms of the vecton theory transform into
 completely standard potentials depending only on scalar fields
 (dilaton, scalaron, other scalars). The scalar formulation  makes it
 easier to look for additional integrals of motion in the
 one-dimensional reductions of DSG.\footnote{
 Pure dilaton gravity is a topological theory and thus  reduces
 to the one-dimensional integrable system. There exist nontrivial
 integrable DGS models involving one massless scalar,  \cite{ATF1}.
 More complex models
 with effectively massive scalar field may have one additional integral,
 at best, and thus generally remain non-integrable. }
 In Section~4 we describe a more standard transition from the
 two-dimensional DVG to unified description of cosmological and static
  solutions in arbitrary gauges and write parametrization invariant Hamiltonian equations of one-dimensional DSG theory. The  canonical  constraint formalism is most convenient in searches for new integrals of motion, which is illustrated by the three-dimensional case. This hints at possible further generalizations of the approach proposed in \cite{ATF1} (pp. 1698-1699), where we derived two nontrivial
  one-dimensional DGS models having two additional
  integrals.

  Section~5 is a central part of this paper. There we introduce the simplest form of the ordinary differential equations describing the generic DGS theory and derive the most important properties of their cosmological and static solutions. In general, this theory is not integrable and we propose several effective approaches to its analytic and qualitative investigation. The simplest one is the power series expansion near the horizons and generalized expansions near the singularities (taking account of nonanalytic terms). We propose a more powerful approach based on an integral equation for the most important function of the system.
  This equation, which we call the \emph{master integral equation},
   summarizes important properties of our
  \emph{cosmo-stat} system.  In completely integrable cases it may be reduced to differential dynamical system. In a few instances, it was possible to, find a global analytic solution of the dynamical system and to draw a picture resembling the classical phase portraits. An example of such a portrait is presented in Appendix.

   In summary, we stress once more that a main goal of this paper is to find approaches to understanding the global picture of cosmological and static solutions of dilaton gravity that couples to nonlinear scalar fields. These theories are not integrable but, possibly, can be approximated by integrable ones. We have shown that there may exist
   additional integrals and have proposed some tools for exploiting this fact. Although main features of the vecton theory hint at its relevance to dark energy, inflation,\footnote{While the relation to dark energy models is evident, the inflation was clearly demonstrated only in numerical studies of the massive vector meson theory in the tachyonic case $m^2 < 0$, \cite{Ford}. }
   and dark matter, the precise relation is still to be uncovered, and the best way to solving this problem is in looking for global qualitative portraits. The difficulty is that the exact system is non-integrable and thus the portrait must be at least three-dimensional.

   Understanding global properties of classical solutions is also desirable for their quantization. The simplest approach was attempted some time ago for classically integrable gravitational systems with minimum number of degrees of freedom (see, e.g.  \cite{CAF},
    \cite{CAF1}, \cite{Henn} and references therein). This primitive quantization can justifiably be criticized for not taking into account space inhomogeneities, which are crucial in cosmological applications  (for a review of more sophisticated ideas in quantum gravity see, e.g., \cite{Hamber}, \cite{QG}). Nevertheless, we hope that such a quantization  of DSG might be of some interest in considering properties of simplest quantum cosmological models.

 \section{Appendix}
 \subsection{Topological portrait}
 Some linear integrals of Section~5 were found in Ref.\cite{ATF1}
 for DGS with $\psi$-independent potentials which, according to (\ref{a.29}),  have the integral $Z(\f)\, \dpsi = \eta_0\,$. If also
 $U_{\xi} = g U$, we find from (\ref{a.34}), (\ref{a.35}) one more integral (recall (\ref{a.27}):
  \be
   Z \dF - g\dvphi \equiv \rho - g \chi = C_1 \,.
  \label{a0.1}
  \ee
  Then it follows that the solution can be expressed in terms of quadratures, see Eqs.(24)-(25) in \cite{ATF1}. An interesting thing not mentioned in \cite{ATF1} is that this integral also exists
 for $\psi$-dependent `multiplicative' potentials
 $U = u(\f) v(\psi)$ when $Z(\f)\, \dpsi$ is not constant.

  Consider the second integrable model of \cite{ATF1}, in which the potential is independent of $\psi$ and depends on an arbitrary
  function $w(\f)$. Then, in Weyl's frame $V$ and $Z$ are given   by
  \be
  U(\xi) \equiv ZV = [\,g_1 \,w(\xi) +\, g_2\, w^{-1} (\xi)\,]^{\pr} \,,\qquad
  [2 g_3 l(\xi) +\, g_4 l^2(\xi)\,] = \xi \,,
   \label{a0.2}
  \ee
   where $d\xi \equiv d\f /Z(\f)$,  $l(\xi) \equiv \ln w(\xi)$, and evidently,
   \be
     l(\xi) = - (g_3/g_4)\biggl[ 1 \mp
     \sqrt{1 + \xi \,(g_4/g_3^2)}\,\biggr]\, \rightarrow \,
     \xi / 2g_3 \,, \quad \textrm{if} \quad g_4 \rightarrow 0\,.
  \label{a0.3}
  \ee
  The third integral was derived by a `brute force' in \cite{ATF1}. In notation of the present paper it is
   \be
  [\rho /\,l^{\pr}(\xi)]^2 - 4g_2 \,h + 2g_4 \,\eta_0^2 \,\ln h =
   C_1 \,.
  \label{a0.4}
  \ee
  The full Lagrangian and the meaning of the coupling constants can be found  in the quoted reference. As shown there, one can derive the solution in quadratures, which are not elementary if $g_4 \neq 0$.
  If we take $g_4 = g_2 = 0$,  we can obtain a simple solution expressing  $w$ as a function of $h$.

  Introducing the scales $w_0$ and $h_0$ for $w(\xi)$  and $h(\xi)$ we find  the relation between normalized $w$ and $h$, which depends only on
  one parameter $\delta$ defined by the relation:
  \be
   2\delta + 1  = \sqrt{1 -8 g_3 \,\eta_0^2 / C_1} \,\geq 0 \,,
  \label{a0.5}
  \ee
   where $g_3 < 0$ for normal scalars and $C_1 >0$.  Then the
   $(h, w)$-portrait of our system is given by
  \be
  w = {|h|^{\delta} \over |1 + \ep |h|^{1+2\delta}| } \,,
  \label{a0.6}
  \ee
 where $\ep \equiv h/|h|$, $-1 < h < \infty $, $0 < w < \infty$.
 Now it is not very difficult to draw the picture of
 the curves describing all possible solutions. In the domain  $h < 0$
 we have static  solutions, while for cosmological ones $h >0$.
 The picture looks like a phase portrait of a dynamical system in the $(h,w)$-plane,
   with  singular points: $(0,1)$, $(1, \half)$, $(0,0)$,
   $(0, \infty)$, $(-1, \infty)$,  $(\infty, 0)$.
 These points are joined by the important separating curves.
 The most interesting points are: the node of the initial singularity,
 $(0,0)$,  the saddle point of the horizon, $(0,1)$,
 and the most interesting `cosmological' point, $(1,\half)$,
 at which all cosmologies tangentially coincide.\footnote{
 Even more interesting but much more complex is the portrait of the model with massless vector (when $g_2 \neq 0$) and scalar, see Eq.(30) in \cite{ATF1}. We will discuss it in a separate publication.}

  This \emph{topological portrait} describes qualitative properties of static and  cosmological solutions for different values of the parameter $\delta$, which characterizes the energy of the massless scalar field. If we deformed it by applying any continuous differentiable transformation  preserving the singularities, it will represent essentially the same cosmo-static system. For example, it is easy to move all singularities to a finite domain of the plane.
  Moreover, to better understand the topological structure of the space of the solutions one may try to extend the portrait to the domains where, say,  $2\delta + 1 < 0$ or $h < -1$, etc. We believe  that the topological portrait is a most adequate global representation of integrable gravitational systems studied in this paper.

 \subsection{Dimensions and Lagrangians}
 The dimensions of geometrical and physical fields and parameters were discussed in \cite{ATFn}. However, in different papers on the affine models we used somewhat different conventions and notation. Here we first give several comments on these matters. In the pure geometrical part, there is no problem at all as we have only the dimension of length. Thus
 $[s_{ij}] = [a_{ij}] = \textrm{L}^{-2}$, $[\bl] = L^0$, geometric Lagrangian (\ref{1.1}) is of dimension $\textrm{L}^{-D}$,
  and geometric action is therefore dimensionless. In transition to a physical picture, we define dimensionless tensor densities
 $\textbf{g}^{ij}$,  $\textbf{f}^{ij}$ by varying Lagrangian (\ref{1.1}). To make the densities dimensionless we multiply it by a constant $2\tga$  where $[\tga] = \textrm{L}^{D - 2}$:
 \be
 \label{a1.1}
  \Lag \equiv 2\tga \sqrt{-\Delta_s}  \,\,\,; \qquad
  {{\p {\Lag}} \over {\p s_{ij}}} \equiv \textbf{g}^{ij} \,, \qquad
 {{\p {\Lag}} \over {\p a_{ij}}} \equiv \textbf{f}^{ij} \, .
  \ee
  Using this definition it is not difficult to prove the following identity ($l$ is a dimensionless constant):
  \be
 \label{a1.2}
  |\Delta_{g}| \equiv \,|\det (\textbf{g}^{ij} +\,
  l \,\textbf{f}^{ij} )| =
  \tga^D\, |\det(s_{ij} +\,
  l^{-1} \,a_{ij})|^{(D-2)/2} \,.
  \ee
  With this identity, one can find the conjugate Lagrangian density
     ${\Lag}^* \, = \, {\Lag}^* (\textbf{g}^{ij} , \textbf{f}^{ij})$:
  \be
 \label{a1.3}
 {\Lag}^* \,=\, \nu^{-1}\,(-\Delta_g /\tga^2)^{\,\nu} \,\,\,;\qquad
 s_{ij} = {{\p {\Lag}^*} \over {\p \textbf{g}^{ij}}} \,, \qquad
 a_{ij} = {{\p {\Lag}^*} \over {\p \textbf{f}^{ij}}} \,.
   \ee
  This result was apparently known to Einstein in the special (four-dimensional) case, when $\nu = 1/2$ and $l = 1$. Following his approach we find the `physical' Lagrangian (\ref{1.2}) as described in \cite{ATF} - \cite{ATFp}. Eqs.(\ref{a1.1}) - (\ref{a1.3}) play the
  key role in this derivation.
  When considering the physical Lagrangian it is better to use the gravitational constant  $\kappa \equiv G/c^4$ to define the physical fields $A_i$ and $F_{ij}$ proportional to $a_i$ and $f_{ij}$ but having the standard dimensions and satisfying the relation
   $[\kappa A_i^2] =1$,
  $[\kappa F_{ij}^2] = \textrm{L}^{-2}$, \cite{ATFn}.
  In this paper we choose the units $c = \kappa =1$ and thus
  $A_i$ is dimensionless and $F_{ij}$ is of dimension
  $\textrm{L}^{-1}\,$. In fact, we use notation $a_i$ for the
  dimensionless potential in this system of units and $f_{ij}$ for the corresponding field tensor $f_{ij} = a_{i,j} - a_{j,i}\,$; this means that $[\la] = \textrm{L}$ and
  $[\la^2 \La] = 1$.
    In Eq.(\ref{1.2}) and in the body of this paper we always apply this
  agreement on notation.
  As we do not use the tensor densities in the main text, the bold-face notation is used only to  $f_{ij} \, f^{ij} \equiv  \textbf{f}^{\,2}\,$    and $a_i \, a^i \equiv  \textbf{a}^{\,2}$.

  One final remark on the dimensions of constants and fields. In our units, we have essentially three dimensional constants of geometric origin: $\La$, $\la$, $\mu \equiv m^2$ ($k_{\nu}$ and $\tga$ are auxiliary dimensional constants having no relation to affine geometry). Coordinates $t, r, u, v$ are of dimension L, the main fields $a_i$, $\psi$, $\f$ are dimensionless
  while $[q] = \textrm{L}^{-1}$. When we denote the scalaron by $\psi$
  (in Section~5) we silently make it dimensionless by including $m$ in its definition, $\psi = q/m$.

  \subsection{Three-dimensional pure vecton theory}
  Let us consider the vecton Lagrangian (\ref{a.2}) in the LC gauge
  (when $\alpha = \gamma$ and $2 \alpha = \ln |h| = F$)
  and without transition to the scalaron description
   \be
     {\cal L}_{\rm v}^{(1)} =
    \lambda_0 \, \f \,\da^2_{\ep} / h \,-\, \dvphi \dF  -
    \mu \,\f \, a^2_{\ep}   \,-\, 2 \Lambda \f h \,,
  \label{a2.1}
  \ee
  where $h = \ep e^{-F}$ and $\lambda_0 \equiv \lambda^2 \Lambda$
  are dimensionless while $\mu = m^2$ is of dimension $L^{-2}$. The Hamiltonian constraint that  must be derived before passing to LC coordinates is
   \be
   {\cal H}_{\rm v}^{(1)} =
    \la_0 \, \f \,\da^2_{\ep} / h \,-\, \dvphi \dF  +
    \mu \,\f \, a^2_{\ep}   \,+\, 2 \La \f h \,=\,0 \,
  \label{a2.2}
  \ee
   while other equations are simply derived by varying Lagrangian (\ref{a2.1}). Instead of the Hamiltonian equations we write
  the equations of motion in the first order form similar
  to (\ref{a.29}). Defining $\xi \equiv (\f^2/2)$,
  $ (\f \,\da) / h \equiv b$, $A \equiv a/\chi$,
  $\mu_1 \equiv \mu/\la_0$, and  denoting $d/d\xi$ by the prime we find:
 \be
  b^{\pr} = -\mu_1 A \,, \quad \prH = \mu A^2 H\,, \qquad
  2\, \xi a^{\pr} = b H \,, \quad
  \xi \prchi  =  (2 \La \xi + \la_0 \,b^2/2) \,H  \,,
    \label{a2.3}
  \ee
 where we omitted index $\ep$ of $a$ and $A$. In the massless limit
 $\mu = \mu_1 = 0$ and therefore $H = H_0$, $b = b_0$ are independent of $\xi$. This immediately gives the three-dimensional Maxwell - Einstein solution. Indeed, we see that
   \be
   h(\xi) = H_0 \,\chi(\xi)\,, \quad
   a = a_0 + (b_0 H_0 /2) \,\ln{\xi/\xi_0} \,, \quad
  \chi = \chi_0 + (\la_0 \,b_0^2 H_0 /2) \,\ln{\xi/\xi_0} \,+\,
  2\La (\xi - \xi_0)\,,
    \label{a2.4}
  \ee
  and thus returning to the variable $\f$ we find (taking $\chi_0 =0$
  and denoting $\la_0 \,b_0^2 H_0 \equiv \tc$):
    \be
    h = H_0 \int_{\xi_0}^{\xi}\,d\xi \biggl({\tc \over 2\xi} \,+\, 2\La \biggr) =\,
    H_0 \int_{\f_0}^{\f}\,d\f \, \biggl({\tc \over \f} \,
    +\, 2\La \f \biggr) \,.
    \label{a2.5}
  \ee
  Equations (\ref{a2.4})-(\ref{a2.5}) define the general solution of
  the pure dilaton gravity with the potential proportional to the expression in the brackets, see Eq.(\ref{a.26}).

  To find approximate solutions for small but finite values of $\mu$, $\mu_1$ we may take $H = H_0$, $b = b_0$, expressions (\ref{a2.4}) for
  $a_0(\xi)$,  $\chi_0(\xi)$, $A_0 = a_0 / \chi_0$ and thus construct successive approximations for $ b_i, H_i, a_i, \chi_i, A_i = a_i/\chi_i$. Our equations (\ref{a2.3}) or their integral formulation are convenient for deriving the  solution near horizons and near singularities at $\xi \rightarrow 0, \infty$.
   In fact, they are simpler than the
   approximate equation studied in \cite{ATF} - \cite{ATFg}.
   At the horizon at $\xi = \xi_0$, when $a_0 = \chi_0 =0$ and
   $H_0, A_0$ are finite, we have the convergent expansions of the solution in powers of $\xi-\xi_0$.
   Moreover, one can see that the radius of convergence cannot be larger than $\xi_0$ (it can be lesser if there is a second horizon). Near singularities we can also find expansions in powers of $\xi$ if we correctly take into account the logarithmic terms seen in the zeroth approximation (\ref{a2.5}).

  A better though much more complex iterations can be derived with the aid of the integral equations for $A(\xi)$ similar to (\ref{a.31}), (\ref{a.32}), if we take the zeroth approximation
   $A =A_0$, solve the equations for $H$ and  $b$,  solve the resulting equation for $A(\xi)$, and then repeat this cycle. A simpler approach is to first find and use an additional approximate integral.
    On this way, the scalaron formulation looks more promising.
   For example, we have seen that the potential for $D=3$ depends on
   $q^2/\xi\,$ and there exists the additional integral (\ref{a.23}),
   which presumably can be used for constructing more effective approximations.

 {\bf Acknowledgment}

 This work was supported in part by the Russian Foundation for Basic Research:
 Grant No. 11-02-01335-a and Grant No.
 11-02-12232-ofi-M-2011.
 Useful remarks of E.A.~Davydov are kindly acknowledged.

 \end{document}